\documentclass[11pt]{article}
\usepackage{graphicx}
\title{Spin Transport in Nanowires}
\author{S. Pramanik and S. Bandyopadhyay\\
Department of Electrical Engineering, Virginia Commonwealth University\\
Richmond, Virginia 23284, USA \\
\\M. Cahay \\
Department of Electrical and Computer Engineering and Computer Science \\
University of Cincinnati \\
Cincinnati, Ohio 45221}
\date{}

\begin{document}

\maketitle

\begin{abstract}
We study high-field spin transport of electrons in a quasi one-dimensional 
channel of a $GaAs$ gate controlled spin interferometer (SPINFET) using a 
semiclassical formalism (spin density matrix evolution coupled with Boltzmann 
transport equation). Spin dephasing (or depolarization) is predominantly 
caused by D'yakonov-Perel' relaxation associated with momentum dependent spin 
orbit coupling effects that arise due to bulk inversion asymmetry (Dresselhaus 
spin orbit coupling) and structural inversion asymmetry (Rashba spin orbit 
coupling). Spin dephasing length in a one dimensional channel has been found 
to be an order of magnitude higher than that in a two dimensional channel. 
This study confirms that the ideal configuration for a SPINFET is one where 
the ferromagnetic source and drain contacts are magnetized along the axis of 
the channel. The spin dephasing length in this case is about $22.5\mu$m at 
lattice temperature of $30$K and $10\mu$m at lattice temperature of $77$K for 
an electric field of $2$kV/cm. Spin dephasing length has been found to be 
weakly dependent on the driving electric field  and strongly dependent on the 
lattice temperature.
\end{abstract}

\section{{Introduction}}
Spin transport in semiconductor nanostructures has attracted significant 
research interest due to its promising role in implementing novel devices 
which operate at decreased power level and enhanced data processing speed. 
Additionally, spin is considered to be the ideal candidate for encoding qubits 
in quantum logic gates \cite{bandy3} because spin coherence time in 
semiconductors \cite{awschalom} is much longer than charge coherence time 
\cite{jariwalla}.

In this paper, we study spin transport of electrons in a quasi one-dimensional 
structure. In the past, we established \cite{pramanik} that in a SPINFET 
configuration where the ferromagnetic source and drain contacts are magnetized 
{\it{along}} the axis of the quasi one-dimensional channel, the spin dephasing 
time ($\tau_s$) is about $3$ ns for an electric field of $2$kV/cm and $10$ ns 
for an electric field of $100$V/cm, at a (lattice) temperature of $30$K. Spin 
dephasing time was much less when the contacts were magnetized along any other 
direction. In the present work, we highlight the spatial variation of spin 
polarization along the channel for this ``large $\tau_s$'' configuration using 
a multi-subband Monte Carlo simulator.

This paper is organized as follows: in the next section, we describe the 
theory followed by a brief description of the Monte Carlo simulator in section 
III and results in section IV. Finally we conclude in section V. 
\section{{Theory}}
\begin{figure}
\centering
\includegraphics[width=4.5in]{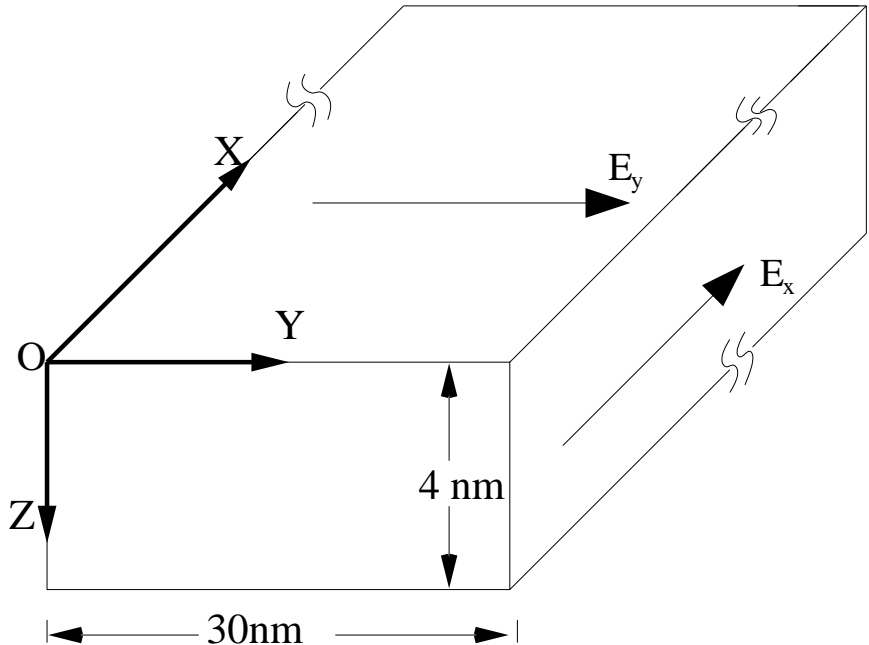}
\caption{Geometry of the nanowire and axis designation (not drawn to scale)}
\label{structure}
\end{figure}

Fig.\ref{structure} shows the schematic of the quasi one-dimensional 
semiconductor structure. An electric field $E_x$ is applied along the axis of 
the quantum wire to induce current flow. In addition, there could be another 
transverse field $E_y$ to cause Rashba spin-orbit interaction. Such a field is 
indeed present in some spintronic devices e.g. a spin interferometer proposed 
in \cite{datta}. Spin polarized electrons are injected at one end of the wire 
from a half-metallic contact with the spin vector oriented along the wire 
axis. Our goal is to investigate how the injected spin polarization decays 
along the channel as the electrons traverse the quantum wire under the 
influence of electric fields $E_x$ and $E_y$ while being subjected to various 
elastic and inelastic scattering events.

In reality, the spin and spatial wavefunctions of the electrons are coupled 
together via {\it{spin-orbit coupling}} Hamiltonian and hence spin dephasing 
(or depolarization) rates are functionals of the electron distribution 
function in momentum space. The distribution function in momentum space 
continuously evolves with time when an electric field is applied to drive 
transport. Thus, the dephasing rate is a dynamic variable that needs to be 
treated self-consistently in step with the dynamic evolution of the electrons' 
momenta. Such situations are best treated by Monte Carlo simulation.

Following Saikin \cite{saikin}, we describe electron's spin by 
standard spin density matrix formalism \cite{blum}:
\begin{eqnarray}
\rho_{\sigma}(t) =
 \left [ \begin{array}{cc}
            \rho\uparrow\uparrow(t) & 
\rho\uparrow\downarrow(t)\\
 \rho\downarrow\uparrow(t)& \rho\downarrow\downarrow(t)\\
             \end{array}   \right ] 
\label{spin_density}
\end{eqnarray}
which is related to the spin polarization component as
\begin{equation}
\label{spin_pol}
S_n(t)=Tr\left(\sigma_n\rho_\sigma(t)\right)
\end{equation}
where $n$ = $x$, $y$, $z$ and $\sigma_n$ denotes Pauli spin matrices. We 
{\it{assume}} that in a small time interval $\delta$t no scattering takes 
place and the electron accelerates very slowly due to the driving electric 
field (in other words, $E_x\delta$t is sufficiently small). During this 
interval $\delta$t we describe electron's transport by a constant ``average 
wavevector'':
\begin{equation}
\label{k}
k=\frac{k_{initial}+k_{final}}{2}=k_{initial}+\frac{qE_x\delta t}{2\hbar}
\end{equation}
During this interval, the spin density matrix undergoes a unitary evolution 
according to
\begin{equation}
\label{evolution}
\rho_\sigma(t+\delta t)=e^{\frac{-iH_{so}(k)\delta 
t}{\hbar}}\rho_\sigma(t)e^{\frac{iH_{so}(k)\delta t}{\hbar}},
\end{equation}
 where $H_{so}(k)$ is the momentum dependent Hamiltonian that has two main 
contributions due to the bulk inversion asymmetry (Dresselhaus interaction) 
\cite{dresselhaus} 
\begin{equation}
\label{dresselhaus}
H_D(k)=-\beta\langle k_y^2\rangle k\sigma_x, (k\equiv k_x)
\end{equation}
and the structural inversion asymmetry (Rashba interaction) \cite{rashba} 
\begin{equation}
\label{rashba}
H_R(k)=-\eta k\sigma_z.
\end{equation}
The constants $\beta$ and $\eta$ depend on the material and, in case of 
$\eta$, also on the external electric field $E_y$ that breaks inversion 
symmetry. Equation (\ref{evolution}) describes a rotation of the average spin 
vector about an effective magnetic field given by the magnitude of the average 
wavevector ($k$) during the time interval $\delta t$. The assumption of 
constant $k$ implies that spin dynamics is coherent  during $\delta t$ and 
there is no dephasing (reduction of magnitude) since the evolution is 
{\it{unitary}}. However, the electric field $E_x$ changes the value of the 
average wavevector $k$ from one interval to the other. Also, the stochastic 
scattering event that takes place between two successive intervals changes the 
value of $k$. These two factors produce a distribution of spin states that 
results in effective dephasing. Thus the evolution of the spin polarization 
vector $\vec S$ (with components $S_x$, $S_y$ and $S_z$) can be viewed as 
coherent motion (rotation) coupled with dephasing/depolarization (reduction in 
magnitude). This type of dephasing is the D'yakonov-Perel' relaxation which is 
a dominant mechanism for spin dephasing in one-dimensional structures.

Another spin dephasing mechanism is the Elliott-Yafet relaxation 
\cite{elliott} which causes instantaneous spin flip during a momentum relaxing 
scattering. However, in quasi one-dimensional structures momentum relaxing 
events are strongly suppressed because of the one-dimensional constriction of 
phase space for scattering. So we can neglect this effect as a first 
approximation. 

The third important spin dephasing mechanism, known as Bir-Aronov-Pikus 
mechanism accrues from exchange coupling between electrons and holes. This 
effect is absent in unipolar transport.

Apart from these three, spin relaxation may take place due to magnetic field 
caused by local magnetic impurities, nuclei spin and other spin orbit coupling 
effects. However, in this work we consider only the D'yakonov-Perel' mechanism 
which is the most important spin dephasing mechanism in quantum wires of 
technologically important semiconductors like $GaAs$.

We can recast (\ref{evolution}) in the following form for the temporal 
evolution of the spin vector:
\begin{equation}
\label{S}
\frac{d\vec S}{dt}=\vec\Omega\times\vec S.
\end{equation}
where the so-called ``precession vector'' $\vec\Omega$ has two components 
$\Omega_D(k)$ and $\Omega_R(k)$ due to the bulk inversion asymmetry 
(Dresselhaus interaction) and the structural inversion asymmetry (Rashba 
interaction) respectively:
\begin{equation}
\label{omegaD}
\Omega_D(k)=\frac{2a_{42}}{\hbar}\left(\frac{\pi}{W_z}\right)^2k
\end{equation}
\begin{equation}
\label{omegaR}
\Omega_R(k)=\frac{2a_{46}}{\hbar}E_yk
\end{equation}
where $a_{42}$ and $a_{46}$ are material constants and $W_z$ is the dimension 
of the wire along $z$ direction. 
In our work we take $a_{42}$ = 2$\times$$10^{-29}$ eV-m$^3$, $a_{46}$ = 
$4$$\times$$10^{-38}$ C-m$^2$ and $E_y$ = 100kV/cm which are reasonable for 
$GaAs-AlGaAs$ heterostructures. 
Solution of (\ref{S}) is straightforward and analytical expressions for 
$S_x(t)$, $S_y(t)$ and $S_z(t)$ can be obtained quite easily (see 
\cite{pramanik}). We note that spin is conserved for every individual electron 
during $\delta t$, i.e.
\begin{equation}
S_x^2(t+\delta t)+S_y^2(t+\delta t)+S_z^2(t+\delta 
t)=S_x^2(t)+S_y^2(t)+S_z^2(t)
\label{conserv}
\end{equation}
\section{{Monte Carlo Simulation}}
The average value of the wavevector in a time interval $\delta t$ depends on 
the initial value of the wavevector $k_{initial}$ at the beginning of the 
interval. Intervals are chosen such that a scattering event can occur only at 
the beginning or end of an interval. The choice of scattering events and the 
wavevector state after the event (i.e. the time evolution of the wavevector) 
are found from a Monte Carlo solution of the Boltzmann transport equation in a 
quantum wire \cite{bandy1}. The following scattering mechanisms are included: 
surface optical phonons, polar and non-polar acoustic phonons and confined 
polar optical phonons. A multi-subband simulation is employed; upto six 
subbands can be occupied in the $y$ direction and only one transverse subband 
is occupied along the $z$ direction even for the highest energy an electron 
can reach. This is because the width of the wire ($y$-dimension) in our 
simulation is $30$ nm and the thickness ($z$ dimension) is only $4$ nm. Hard 
wall boundary conditions are applied. The details of the simulator can be 
found in \cite{bandy1}.

We solve (\ref{S}) directly in the Monte Carlo simulator. The simulation has 
been carried out in the absence of any external magnetic field, although the 
Monte Carlo simulation allows inclusion of such a field.

\section{{Results and Discussion}}
We consider the case when the electrons are injected at the left end ($x$ = 
$0$) of the quantum wire  with their spins initially polarized along the axis 
of the wire (i.e along $x$ axis). In order to maintain current continuity, as 
soon as an electron exits from the right end of the wire, it is re-injected at 
the left end. The spin polarization of this electron is re-oriented along $x$ 
direction.
\subsection{{Effect of driving electric field:}}
\begin{figure}
\centering
\includegraphics[width=6.8in]{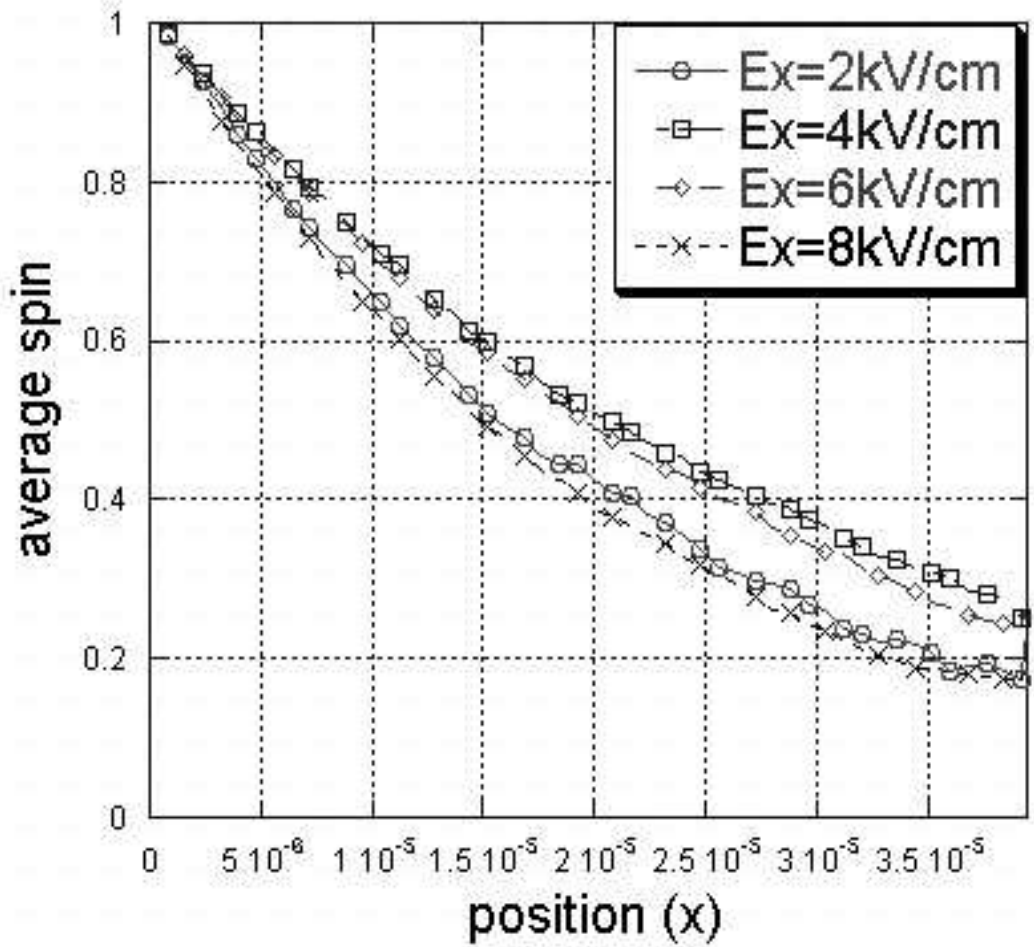}
\caption{Spatial dephasing of ensemble average spin vector in $GaAs$ quantum 
wire at $30$K for various driving electric fields. Spins are injected with 
their polarization initially aligned along the wire axis. Position x is 
measured in meters.}
\label{compare_E}
\end{figure}

In Fig.\ref{compare_E}, we show how the magnitude of the ``average spin 
vector''$\langle S\rangle$ (defined later) decays along the channel for four 
different values of the driving electric field $E_x$. For this study, we have 
set lattice temperature = $30$K and length of the channel = $40\mu$m. 

The quantity $\langle S\rangle$ is calculated as follows:
at each point $x$ along the channel we compute ``ensemble average'' of the 
spin components during the entire evolution time $T$. Mathematically, this can 
be expressed as 
\begin{equation}
\label{average_i}
\langle S_i\rangle(x,T)=\frac{\sum_{t=0}^{T}\sum_{n=1}^{n_x(x,t)} 
S_{i,n}(t)}{\sum_{t=0}^{T}n_x(x,t)}.
\end{equation}
Here $i$ denotes the components $x$, $y$ and $z$, $n_x(x,t)$ denotes number of 
electrons in a bin of size $\delta x$ centered around $x$ at time $t$, 
$S_{i,n}(t)$ denotes the $i$ th component of spin corresponding to the n th 
electron and is derived from equation (\ref{S}). The ``average spin vector'' 
is defined as:
\begin{equation}
\label{average}
\langle S\rangle (x,T)= \left[\langle S_x\rangle^2+ \langle S_y\rangle^2+ 
\langle S_z\rangle^2\right]^\frac{1}{2}
\end{equation}

Fig.\ref{compare_E} is the snapshot of this ``time-averaged ensemble spin 
polarization'' ($\langle S\rangle$) along the channel for $T$ = $5$ ns. It has 
been found (though not shown) that $\langle S\rangle$ is almost independent of 
the evolution time $T$ for $T\geq5$ns. So we can deem Fig.\ref{compare_E} as 
``steady state'' spin polarization along the channel.

We note that the decay characteristic resembles an exponential trend. 
Therefore we define the spin dephasing length ($l_s$) as the distance 
(measured from left end) where $\langle S\rangle$ decays to $\frac{1}{e}$ 
times its initial value of $1$. For $E_x$ = $2$ kV/cm (or voltage across the 
channel = $V_{DS}$ = $8$ volts), $l_s\sim$ $22.5$$\mu$m.

Earlier, spin transport of electrons in III-V semiconductor quantum well and 
heterostructures was carried out by Bournel et al. \cite{bournel} and Privman 
et al. \cite{saikin, saikin1}. Typical spin dephasing length was found to be 
$\sim$  $1\mu$m. In our present study we observe {\it{an order of magnitude 
increase}} in spin dephasing length. This is due to suppression of momentum 
relaxing events in one dimension (i.e. electron-phonon scattering) that also 
suppresses D'yakonov-Perel' spin relaxation.

We also note from Fig.\ref{compare_E} that spin polarization along the channel 
is weakly dependent on driving electric field $E_x$. At higher applied voltage 
(e.g. $E_x$ = $4$kV/cm), drift velocity is larger compared to the case when 
$E_x$ = $2$ kV/cm. This effect leads to slight increase in the spin dephasing 
length. Also at high electric field, scattering probability increases which 
tends to decrease the spin dephasing length. We observe that for $E_x$ = 
$6$kV/cm or $8$kV/cm the later effect dominates over the first, leading to 
smaller $l_s$.
\clearpage
\clearpage
\subsection{{Decay of spin components:}}
\begin{figure}
\centering
\includegraphics[width=6in]{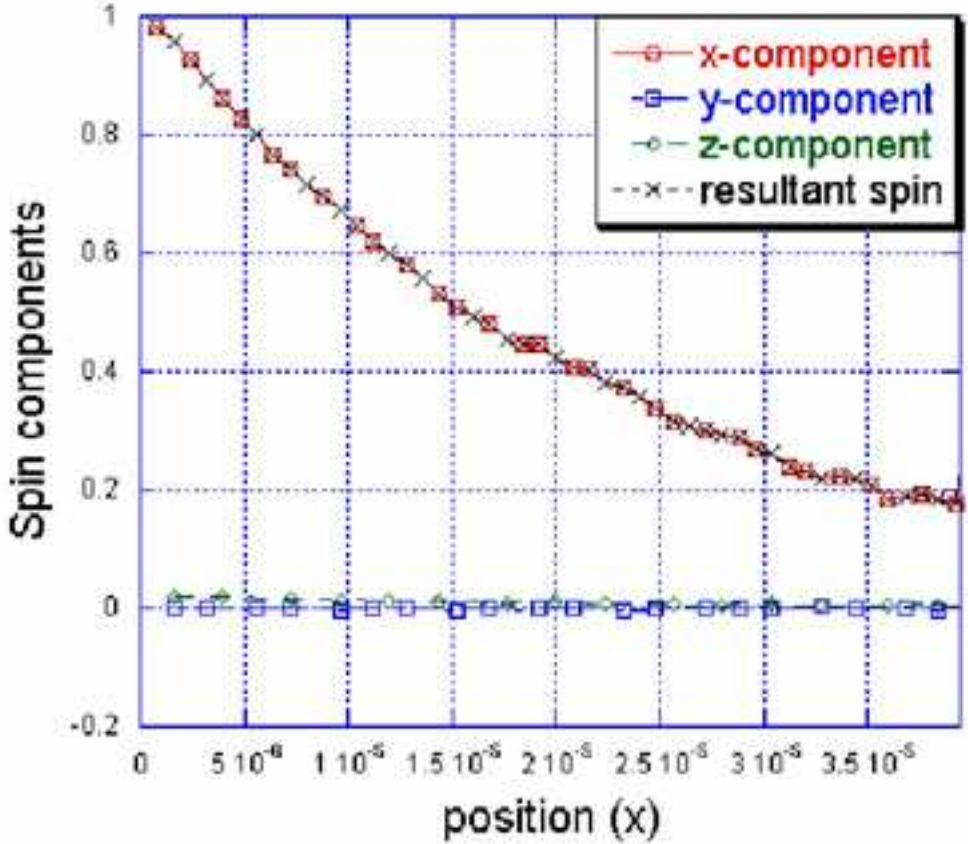}
\caption{Spatial dephasing of the $x$, $y$ and $z$ components of spin in the 
$GaAs$ quantum wire structure at $30$K. The driving electric field ($E_x$) is 
$2$kV/cm and the spins are injected with their polarization initially aligned 
along the wire axis. Position x is measured in meters.}
\label{components}
\end{figure}
Fig.\ref{components} shows how the average spin components (defined as 
$\langle S_i\rangle$ in (\ref{average_i})) decay along the channel. The 
driving electric field $E_x$ = $2$kV/cm and the lattice temperature is $30$K. 
Since, initially the spin is polarized along the $x$ direction, the ensemble 
averaged $y$ and $z$ components remain near zero and the ensemble averaged $x$ 
component decays along the channel. The decay of the ensemble averaged $x$ 
component is indistinguishable from the decay of $\langle S\rangle$ (defined 
in (\ref{average})) shown in Fig.\ref{components}.

It is interesting to note that the spatial decay of the $x$ component along 
the channel is monotonic with no hint of any oscillatory component which 
generally manifests for $y$ and $z$ polarized injections (not shown in this 
paper but see \cite{pramanik}). The oscillatory component is a manifestation 
of the coherent dynamics (spin rotation) while the monotonic decay is a result 
of incoherent dynamics (spin dephasing or depolarization). There is a 
competition between these two dynamics determined by the relative magnitudes 
of the rotation rate ($\Omega$) and the dephasing rate. For the $x$ component, 
the rotation rate is weak because it is solely due to Rashba interaction which 
is weak. Hence the dephasing dynamics wins handsomely resulting in no 
oscillatory component.

\section{{Conclusion}}
In this paper, we have shown how spin dephases in a quasi one dimensional 
structure. It has been found that the spin dephasing length is $\sim$ $10\mu$m 
in a $GaAs$ $1$-D channel at liquid nitrogen temperature which is an 
{\it{order of magnitude}} improvement over two dimensional channels. This 
effect can be exploited in the design of a gate controlled spin interferometer 
where the suppression of spin dephasing is a critical issue.

\bigskip

\noindent {\bf Acknowledgement}: This work is supported by the National
Science Foundation.


\begin{thebibliography}{10}

\bibitem{bandy1}
Telang, N. and Bandyopadhyay, S.,
"Effects of a Magnetic Field on hot electron transport in quantum wires",
Applied Physics Letters, (1995), {\bf 66}, 1623--1625.

\bibitem{awschalom}
Kikkawa, J.M. and Awschalom, D.D., "Resonant spin amplification in n-type 
${G}a{A}s$", Physical Review Letters, (1998), {\bf 80}, 4313--4316.

\bibitem{bandy3}
Bandyopadhyay, S., "Self assembled nanoelectronic quantum computer based on 
the {R}ashba effect in quantum dots", Physical Review B, (2000), {\bf 61}, 
13813--13820.

\bibitem{jariwalla} 
Mohanty, P., Jariwalla, J.M.Q. and Webb, R.A.,
"Intrinsic decoherence in mesoscopic systems", Physical Review Letters,
(1997), {\bf 78}, 3366--3369.

\bibitem{bandy2}
Telang, N. and Bandyopadhyay, S., "Hot-electron magnetotransport in quantum 
wires", Physical Review B, (1995), {\bf 51}, 9728--9734.

\bibitem{blum}
Blum, K., {\it Density matrix theory and applications}, (Plenum Press, 1996,
New York, 2nd. edition).


\bibitem{datta}
Datta, S. and Das, B., "Electronic analog of the electro-optic modulator",
Applied Physics Letters, (1990), {\bf 56}, 665--667.

\bibitem{dresselhaus}
Dresselhaus, G., "Spin-orbit coupling effects in zinc blende structures",
Physical Review, (1955),
580--586.

\bibitem{elliott}
Elliott, R.J., "Theory of the effect of spin-orbit coupling on
		  magnetic resonance in some semiconductors", Physical Review,
(1954), 266-279.



\bibitem{leburton}
Jovanovich, D. and Leburton, J.P., in
{\it Monte Carlo device simulation: full band and beyond},
(Kluwer Academic, 1991, Boston).

\bibitem{rashba}
Bychkov, Y.A. and Rashba, E.I.,
"Oscillatory effects and the magnetic susceptibility
		  of carriers in inversion layers",
Journal of Physics C: Solid State Physics",
(1984), {\bf 17}, 6039--6045.

\bibitem{saikin}
Saikin, S. and Shen, M. and Cheng, M.C. and Privman,
		  V., "Semiclassical Monte Carlo Model for in-plane
		  transport of spin-polarized electrons in {III}-{V} 
heterostructures.", www.arXiv.org/cond-mat/0212610, (2002).

\bibitem{saikin1}
Shen, M. and Saikin, S. and Cheng, M.C. and Privman,
		  V., "Monte Carlo simulation of spin polarized transport",
www.arxiv.org/cond-mat/030239 (2003).

\bibitem{pramanik}
Pramanik, S. and Bandyopadhyay, S. and Cahay, M., "Spin dephasing in quantum 
wires", Physical Review B (in review).

\bibitem{bournel}
Bournel, A. and Dollfus, P. and Bruno, P. and Hesto,
		  P.,
"Spin polarized transport in 1-D and 2-D
		  semiconductor heterostructures",
Materials Science Forum, (1999), {\bf 297--298}, 205--212.

\end{thebibliography}
\end{document}